\RequirePackage{lineno}
\documentclass[aps,pre,preprint,superscriptaddress,showpacs]{revtex4-1}
\usepackage{amssymb}
\usepackage{epsfig}
\usepackage{amsmath}
\usepackage{times}
\usepackage{color}
\usepackage{lipsum}
\usepackage{ulem}
\usepackage{sidecap}

\begin{document}
\title{\Large{Spiral wave chimeras in populations of oscillators coupled to a slowly varying diffusive environment}}
\author{Lei Yang}
\author{Yuan He}
\affiliation{School of Physics, Hangzhou Normal University,
Hangzhou 311121, China}
\author{Bing-Wei Li}
\email[Corresponding author. Email address: ]{bwli@hznu.edu.cn}
\affiliation{School of Physics, Hangzhou Normal University, Hangzhou 311121, China}
\begin{abstract} 
Chimera states are firstly discovered in nonlocally coupled oscillator systems. Such a nonlocal coupling arises typically as oscillators are coupled via an external environment whose characteristic time scale $\tau$ is so small (i.e., $\tau \rightarrow 0$) that it could be eliminated adiabatically. Nevertheless, whether the chimera states still exist in the opposite situation (i.e., $\tau \gg 1$) is unknown. Here, by coupling large populations of Stuart-Landau oscillators to a diffusive environment,  we demonstrate that spiral wave chimeras do exist in this oscillator-environment coupling system even when $\tau$ is very large. Various transitions such as from spiral wave chimeras to spiral waves or unstable spiral wave chimeras as functions of the system parameters are explored. A physical picture for explaining the formation of spiral wave chimeras is also provided. The existence of spiral wave chimeras is further confirmed in ensembles of FitzHugh-Nagumo oscillators with the similar oscillator-environment coupling mechanism. Our results provide an affirmative answer to the observation of spiral wave chimeras in populations of oscillators mediated via a slowly changing environment and give important hints to generate chimera patterns in both laboratory and realistic chemical or biological systems.
\end{abstract}
\date{\today }
\maketitle

\section{Introduction}

The coexistence of synchronized and desynchronized states in populations of identically coupled oscillators is known as a chimera state \cite{kuramoto_npcs02,abrams_prl04}. Since its first discovery by Kuramoto and Battogtokh in 2002 \cite{kuramoto_npcs02}, this counterintuitive and exotic state has inspired massive theoretical \cite{kuramoto_npcs02,abrams_prl04,omelchenko_prl08,sethia_prl08,abrams_prl08,omelchenko_prl11,omelchenko_prl13,sethia_pre13,zakharova_prl14,xu_prl18,zhang_prx20,zhu_pre14,zheng_sspma20,zhang_prl21,dai_fop20} and experimental activity \cite{hagerstrom_nat12,tinsley_nat12,nkomo_prl13,martens_pnas13,wickramasinghe_pccp14, schmidt_cha14,gambuzza_pre14,larger_nc15}. For detailed reviews on chimera states, see Refs \cite{panaggio_non15,omelchenko_non18,parastesh_pr21}. In the theoretical studies, for example, chimera states have been observed in a variety of coupled oscillator systems whose intrinsic dynamics of the element could be described by phase oscillators \cite{kuramoto_npcs02,abrams_prl04}, limit-cycle oscillators \cite{omelchenko_prl13}, excitable oscillators \cite{semenova_prl16,dai_nd18} and chaotic maps \cite{omelchenko_prl11}. In the experimental studies, chimera states have been realized in a spatial light modulator feedback system \cite{hagerstrom_nat12}, chemical Belousov-Zhabotinksy (BZ) oscillators \cite{tinsley_nat12,nkomo_prl13}, mechanically coupled metronomes \cite{martens_pnas13}, electrochemical systems \cite{wickramasinghe_pccp14,schmidt_cha14}, electronic oscillators \cite{gambuzza_pre14}, and lasers \cite{larger_nc15}, to name a few. With these studies, the conditions of generating chimera states have been largely relaxed compared to those adopted in the seminal works. Particularly, it is believed for a long time that chimera states occur only in the nonlocal coupled systems \cite{kuramoto_npcs02,abrams_prl04}, but recent works show that the nonlocal coupling is not necessary at all and chimera states can be observed in systems with different coupling topologies \cite{bera_epl17}, including global coupling \cite{sethia_prl14,yeldesbay_prl14} and local couping \cite{laing_pre15,clerc_pre16,bera_pre16_2,premalatha_chaos18,Kundu_pre18,clerc_cha18}. Further more, the concept of chimera states has been broaden and generalized largely. In addition to classical chimera states, various new types of chimera states have been observed such as clustered chimeras \cite{sethia_prl08}, chimera death \cite{zakharova_prl14}, amplitude-mediated chimeras \cite{sethia_pre13} and amplitude chimera \cite{zakharova_prl14}, alternating chimeras \cite{haugland_sr15}, traveling chimeras \cite{xie_pre14}, spiral wave (scroll wave) chimeras in two (three) dimensional systems \cite{kuramoto_ptps03,shima_pre04,martens_prl10}. Besides the theoretical and experimental interest, chimera states may also have biological implications \cite{rattenborg_nbr00,tamaki_cb16,lainscsek_cha19,huo_nsr21,majhi_plr19,wu_fop22}. For example, for some birds and many mammal animals such as dolphin, one part of the brain keeps active (synchronized), while the other part of the brain remains inactive (desynchronized) when they sleep \cite{rattenborg_nbr00}, which is well known as the unihemispheric slow-wave sleep. What is more, the chimera state may also be related to the first-night effect in human sleep, where one hemisphere is more vigilant than the other due to an unfamiliar environment during sleep \cite{tamaki_cb16}. Recently, it has been shown that chimera states may also exist in the human brain under some conditions such as epileptic seizures \cite{lainscsek_cha19}.

Chimera states are observed not only in one dimensional systems \cite{kuramoto_npcs02} but also in two or three dimensional systems  \cite{shima_pre04,martens_prl10,laing_phd09,gu_prl13,tang_jcp14,xie_pre15,maistrenko_njp15,li_pre16,kundu_epj18,guo_CSF18,rybalova_chaos19,totz_np18,totz_sr20,li_cnns21,bera_cha21,bataille_pre21,tian_fop17}. One of the most remarkable examples in the two dimensional system is the so-called spiral wave chimera (SWC) which combines the features of spiral waves and chimera states \cite{kuramoto_ptps03,shima_pre04,martens_prl10,laing_phd09,gu_prl13,tang_jcp14,xie_pre15,li_pre16,kundu_epj18,guo_CSF18,rybalova_chaos19,totz_np18,totz_sr20,bataille_pre21,li_cnns21}. Differing from the classical spiral wave whose core center is a phase singularity (or topological defect) at which the amplitude drops to zero, the core region of the SWC consists of a group of desynchronized oscillators running at full amplitude \cite{kuramoto_ptps03,shima_pre04,martens_prl10}. Since the first discovery of such a state in the nonlocally coupled oscillator system by Kuramoto and Shima \cite{kuramoto_ptps03,shima_pre04}, SWCs have received growing interests in the last decade. In the theoretical studies, an analytical description of such a SWC is provided in the nonlocally coupled phase model, and by the perturbation method the size of the incoherent core and rotating speed could be predicted \cite{martens_prl10}. Besides periodic oscillators, SWCs have also been reported in a complex and chaotic oscillator system \cite{gu_prl13}. It is worth mentioning that instead of nonlocally coupled systems,  Li {\it et al}. showed the existence of SWCs in locally coupled reaction-diffusion (RD) systems\cite{li_pre16}. Recently, Totz {\it et al.} reported the first experimental verification of the existence of spiral wave chimeras using large populations of nonlocally coupled BZ chemical oscillators and explored the transition from stable SWCs to unstable ones \cite{totz_np18}.

For the aforementioned systems where chimera states are observed involving either nonlocal, local or global couplings, the interactions among oscillators are assumed to be direct. In many physical and biological systems, however, the individuals do not interact directly but rather by means of a common environment \cite{camilli_sci06,garcia_pnas04,monte_pnas07,toth_jpcb06,taylor_sci09,gregor_sci10,danino_nat10,shutz_BJ11,noorbakhsh_pre15,rubin_prl13,gou_jns16,iyaniwura_ads21}. A well-known example is related to some bacteria. For them, the individuals communicate with each other through signaling molecules that are released into the extracellular environment, and dynamical quorum sensing (QS) occurs once the population density is beyond a critical value \cite{camilli_sci06,monte_pnas07,taylor_sci09}. Other examples involving environmental couplings include genetic oscillators \cite{garcia_pnas04,monte_pnas07},  BZ chemical oscillators \cite{taylor_sci09,toth_jpcb06}, slime mold {\it Dictyostelium discoideum} \cite{gregor_sci10,noorbakhsh_pre15}, yeast cells \cite{shutz_BJ11}, and neural oscillators \cite{rubin_prl13}. 

It is often assumed that the external environment through which the individuals communicate is well-stirred \cite{camilli_sci06,garcia_pnas04,monte_pnas07,toth_jpcb06,taylor_sci09,gregor_sci10}. Nevertheless, there is also a growing evidence that diffusion effects of chemical signaling molecules in the extracellular medium should be considered, which usually leads to a RD model representing a population of oscillators coupled via a diffusive environment \cite{danino_nat10,shutz_BJ11,noorbakhsh_pre15}. Synchrony and various spatiotemporal patterns have been reported in this kind of oscillator-environment coupling (OEC) systems \cite{chandrasekar_pre16,choe_pre20}. In fact, the OEC system is the first model showing the existence of the SWC \cite{shima_pre04}. In the seminal work of Kuramoto {\it et al}., to observe chimera states in the FitzHugh-Nagumo (FHN) OEC system, the time scale of the dynamical environment is assumed to be zero. But it is found later that this assumption seems unnecessary to observe chimera states \cite{li_pre16,laing_pre15}.

As noticed by Kuramoto and his colleague \cite{shima_pre04}, the nonlocal coupling could be derived from OEC systems by eliminating adiabatically the environment component. In realistic systems such as biological systems, however, it is also quite common that the time scale of the external environment may be much larger than their intrinsic time scale \cite{rubin_prl13}. For this case, the external environment through which the oscillators are coupled changes slowly compared to the oscillation of the oscillators. Under such a condition, adiabatic approximation which leads to the spatially nonlocal coupling breaks down. As the nonlocal coupling plays a vital role in generating chimera states, an interesting question naturally arise: whether chimera states could be generated in this opposite case?

In this work, we first consider a system representing large populations of Stuart-Landau oscillators coupled to a slowly changing diffusive environment and show the existence of SWCs in this system. For this model, it is found that SWCs occur for $K'_{c}<K<K_{c}$ where $K$ denotes the coupling strength and $K_{c}$ and $K'_{c}$ are two critical values. If $K>K_{c}$, the desynchronized core vanishes and SWCs become spiral waves; while $K<K'_{c}$, SWCs would become unstable. The reversal of rotation direction of SWCs as we change the sign of the system parameter is found and analyzed. The emergence of SWCs can be explained from the point view of synchronization driven by the periodic forcing. Furthermore, the existence of SWCs is also confirmed in populations of FHN oscillators mediated via a slowly changing environment. Our current results together with previous findings suggest that OEC systems may be a kind of the universal system to observe SWCs. These findings provide key hints to explore the chimera states in laboratory and realistic chemical and biological systems.

\section{Spiral wave chimeras in an environmentally coupled Stuart-Landau oscillator system}

\subsection{The Stuart-Landau oscillator model with the environmental coupling}
A general model that represents a large population of oscillators coupled via a diffusive environment usually reads \cite{noorbakhsh_pre15,danino_nat10,li_jnls17,cao_chaos19},
\begin{eqnarray}
\frac{\partial\mathbf{Z}}{\partial t} &=& \textbf{F}(\mathbf{Z})+\mathbf{H}(\mathbf{Z},\mathbf{z}), \label{gene_eq1} \\
\tau \frac{\partial\mathbf{z}}{\partial t} &=& \mathbf{G}(\mathbf{Z},\mathbf{z}) + \mathbf{D}_{\mathbf{z}}\nabla^2\mathbf{z}.\label{gene_eq2}
\end{eqnarray}
The column vectors $\mathbf{Z}(\mathbf{r},t)$ and $\mathbf{z}(\mathbf{r},t)$ represent the dynamical state of the oscillator located at the position $\mathbf{r}$ and the external environment that the oscillator senses, respectively. The intrinsic dynamics of the oscillator is governed by $\partial_{t} \mathbf{Z}=\mathbf{F}(\mathbf{Z})$. The functions $\mathbf{H}(\mathbf{Z},\mathbf{z})$ and $\mathbf{G}(\mathbf{Z},\mathbf{z})$ are the interaction terms, which denote the effects of the environment on the oscillators and the effects of oscillators on the environment, respectively. The parameter $\tau$ represents the relative time scale of $\mathbf{z}$ to $\mathbf{Z}$. The term $\mathbf{D}_{\mathbf{z}}\nabla^2\mathbf{z}$ in Eq. (\ref{gene_eq2}) is added to account for the diffusion of signaling molecules in the external environment with the diffusion constant (matrix) $\mathbf{D}_{\mathbf{z}}$.

For the specific model, we take the Stuart-Landau (SL) oscillator as the local dynamics and linear interaction between the oscillator and the environment is further assumed. Explicitly, the model we are going to study could be written as
\begin{eqnarray}
\frac{\partial W}{\partial t} &=& W-(1+i\alpha)|W|^2W+K(S-W), \label{SL1}\\
\tau_{s} \frac{\partial S}{\partial t} &=& W-S+D_{s}\nabla^2S. \label{SL2}
\end{eqnarray}
Here $W(\mathbf{r},t)$ is a space-time dependent complex variable representing the state of the SL oscillator and $S$ is a complex-valued diffusive field denoting the external environment. Compared to Eqs. (\ref{gene_eq1}) and (\ref{gene_eq2}), one finds that $\textbf{F}(W)= W-(1+i\alpha)|W|^2W$ where $\alpha$ is the intrinsic frequency of the oscillator, and the interaction terms $\mathbf{H}(W,S)=K(S-W)$ and $\mathbf{G}(W,S)=W-S$ with $K$ being the coupling strength between the oscillators and the environment. Clearly, the above system represents a large population of SL oscillators coupled via a diffusive environment $S$. This kind of system resembles the physical model proposed to study pattern formation in the BZ reaction dispersed in water droplets of a water-in-oil
aerosol OT (AOT) microemulsion system (BZ-AOT system) \cite{alonso_jcp11} and to model spot dynamics in gas discharges \cite{schenk_prl97}.

Note that if the environment changes extremely fast, i.e., $\tau_{s}\rightarrow0$, and then Eq. (\ref{SL2}) can be solved using the Green function approach. Consequently, Eqs. (\ref{SL1}) and (\ref{SL2}) are reduced to a nonlocal coupling system which likes \cite{kuramoto_ptps03,shima_pre04}  
\begin{eqnarray}
\frac{\partial W}{\partial t} &=& W-(1+i\alpha)|W|^2W+K\int{G(\mathbf{r'},\mathbf{r})(W(\mathbf{r'},t)-W(\mathbf{r},t))d^{2}\mathbf{r'}}, \label{SL_nonlocal}
\end{eqnarray}
where $G(\mathbf{r'},\mathbf{r})$ is the core of the Green function \cite{kuramoto_ptps03,shima_pre04}. The above case with the nonlocal coupling has been extensively considered in last decades. However, there is very few work on chimera states in the opposite limit, i.e., $\tau_{s} \gg 1$, which means the environment is inertial or evolves slowly ($|\partial S/\partial t|\ll 1$).

\subsection{Numerical methods and measurements} 

We employ the fourth Runger-Kutta method to solve these coupled equations with a space step $dx=dy=0.2$ and a time step $dt=(dx)^2/5=0.008$. The system is composed of $N \times N$ grid points with $N=1024$. To generate a SWC, the cross-field initial condition and no-flux boundary condition are used. As we consider the situation of the slowly changing field, we set $\tau_{s}=100$. The effective diffusion constant is chosen as $D_{\rm eff} = D_{s}/\tau_{s} =1$ for simplicity. The other parameters such as $\alpha$ and the coupling strength $K$ are taken as control parameters, and we want to see how the dynamics of SWCs emerge and change as such parameters vary.

To quantify the size of the incoherent core of SWCs through our work, we introduce an time-averaged order parameter,
\begin{eqnarray}
\left<{\rm R}_{j,k}\right> = \frac{1}{2d+1}\left<\left|\sum_{\left<j,k\right>}e^{{\rm i}\phi_{j,k}}\right|\right>
\end{eqnarray}
where $\phi_{j,k}=\tan^{-1}({\rm Im}W/{\rm Re}W)$ denotes the oscillation phase in the complex plane of $W$ and $\left<\cdot\right>$ means the average over a certain time interval (e.g., $\Delta T=5,000$ in this work). The notation $\left<j,k\right>$ means the set of the nearest neighbor oscillators including itself and $1/(2d+1)$ is a normalization factor with $d$ is the number of nearest oscillators along the one dimension ($d=2$ here). From the definition of the order parameter, it is straightforward to see that for the coherent region $\left<{\rm R}\right> \approx 1$, while in the incoherent region, $\left<{\rm R}\right>$ should be less than unit. For a SWC, we will see that there is a circular region with $\left<{\rm R}\right> < 1$ for the desynchronized core and we then could measure its diameter, $d_{core}$, which is defined as $d_{core}=(d_{x}+d_{y})/2$ where $d_{x}(d_{y})$ means the distance between grid points along the center $x(y)$ line when $\left<R\right>\le  0.98$.

\subsection{Existence and characterization of spiral wave chimeras}

\begin{figure}[t]
\includegraphics[width=\linewidth]{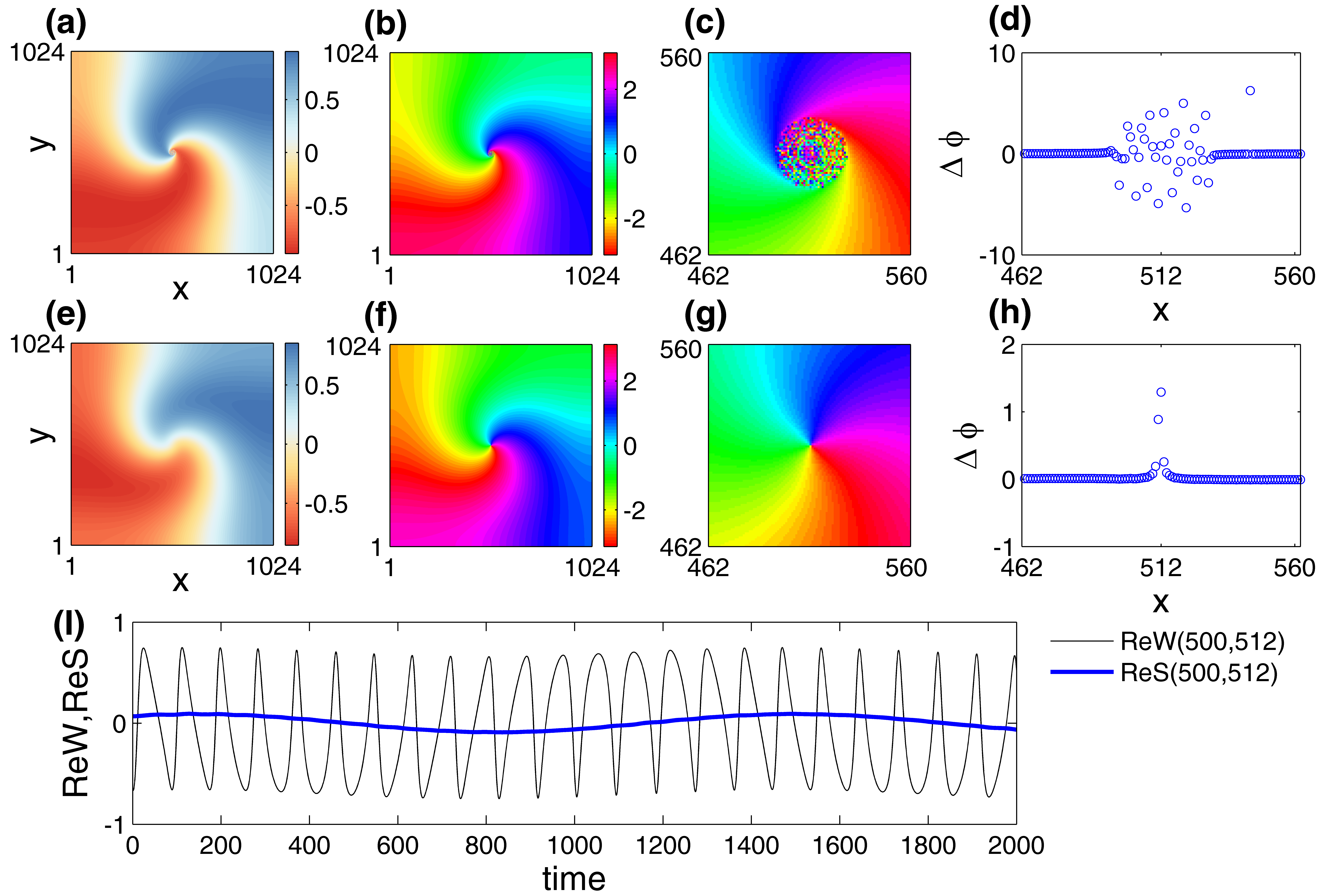}	
\caption{(color online). SWCs in populations of SL oscillators coupled via a slowly changing environment. (a) A SWC for the real component of $W$, and (b) the corresponding phase, i.e., $\phi$, of SWCs, and (c) enlarged view of the core region in (b). (d) The variation of $\Delta \phi = \phi_{i+1,N/2}-\phi_{i,N/2}$ with respect to $x$ along the horizontal central axis in (c). (e) A spiral wave for the real component of $S$, and (f) the corresponding phase of spiral waves and (c) enlarged view of the core region in (f). (h) The variation of $\Delta \phi$ with respect to $x$ along the horizontal central axis in (g). (I) Temporal profile of ${\rm Re} W_{i,j}$ and ${\rm Re}S_{i,j}$ inside the core region with $i=500$ and $j=512$. Parameters are $\alpha=-0.2$ and $K=0.5$.}\label{swc}
\end{figure}

\begin{figure}[t]
\includegraphics[width=\linewidth]{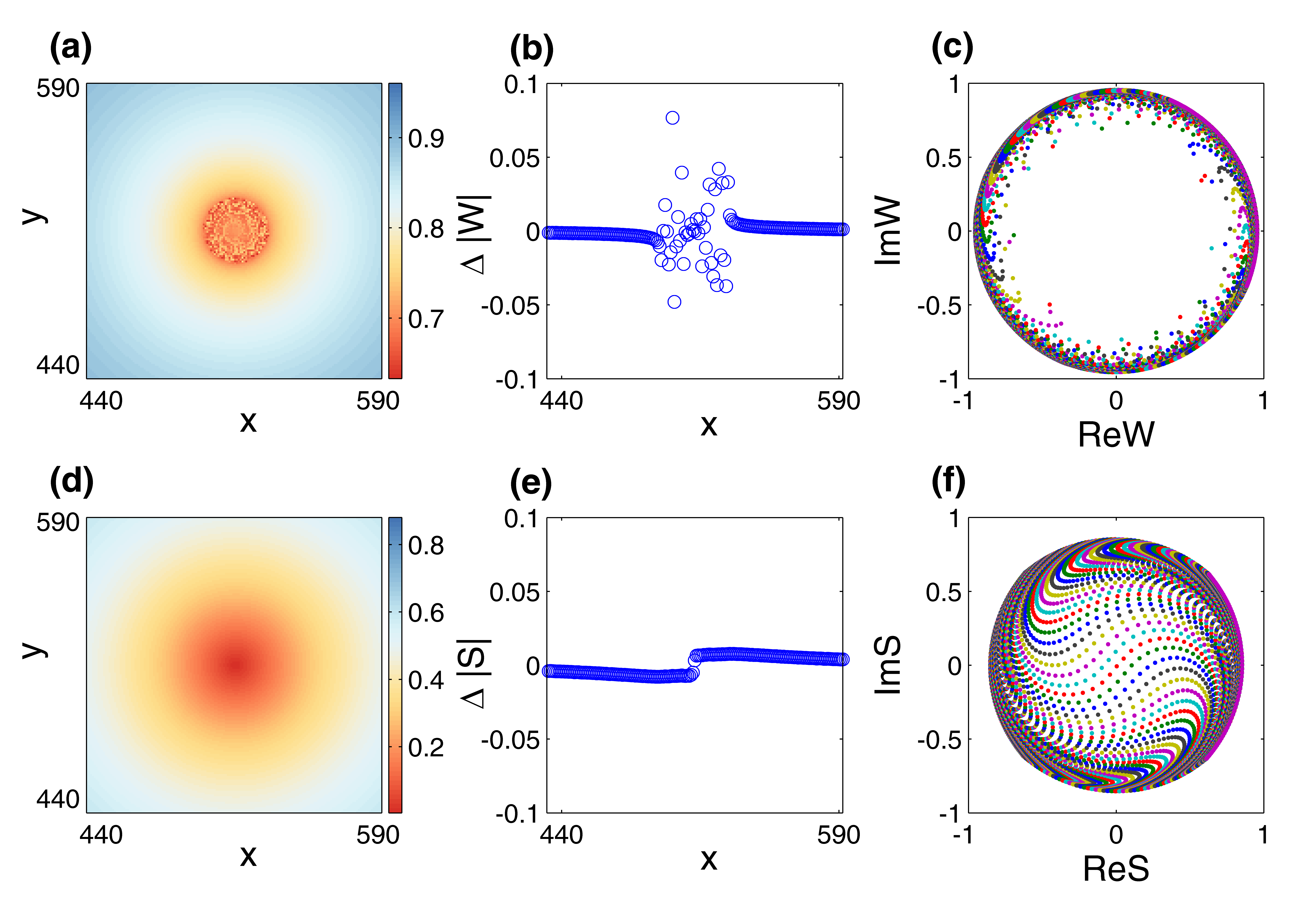}
\caption{(color online). The amplitude (modules) of the component of $W$ and $S$ and corresponding phase portraits. (a) $|W|$ near the core and (b) the variation of $\Delta |W|=|W_{i+1,N/2}|-|W_{i,N/2}|$ with respect to $x$ along the center line. (c) The phase portrait in the ${\rm Re}W-{\rm Im}W$ plane. (d) $|S|$ near the core and (e) the variation of $\Delta |S|=|S_{i+1,N/2}|-|S_{i,N/2}|$ with respect to $x$ along the center line. (f) The phase portrait in the ${\rm Re}S-{\rm Im}S$ plane. All the parameters are the same as in Fig.\ref{swc}.}\label{swc_amp}
\end{figure}

Figure \ref{swc} shows a typical SWC pattern in populations of SL oscillators coupled via a diffusive field for $K=0.5$ and $\alpha=-0.2$. The snapshot of ${\rm Re} W$ (real part of the complex variable $W$) , the phase $\phi$ and its enlarged view of the core region are shown in Fig. \ref{swc}(a)-(c). From these panels, one finds that oscillators in the spiral arm are phase-locked and show continuous behaviors in space, while in the circular-shaped core region they behave desynchronized in time and seem spatially discontinuous. This discontinuous property can be further demonstrated from Fig. \ref{swc} (d) showing the $\Delta \phi = \phi_{i+1,N/2}-\phi_{i,N/2}$ along the  center horizontal line. As we expect, the fluctuation of $\Delta \phi$ in the core region is quite large but almost vanishes in the region away from the core. Due to the diffusion, the environment variable $S$ exhibits a normal spiral pattern, i.e., the core region is smooth and continuous as shown in Fig. \ref{swc}(e-g). Different from the component $W$, a well defined phase singularity for the variable $S$ can be identified which is approximately at the center of the system. It is thus that a big fluctuation only happens in the phase singularity, and except that the distribution of $\phi$ is smooth as shown in Fig. \ref{swc}(h). The typical temporal profile of ${\rm Re}W$ and ${\rm Re}S$ inside the core region is shown in Fig. \ref{swc}(I) and one finds that in this case $S$ evolves slower than $W$.

The SWC reported here is an amplitude-mediated chimera state. That is, in addition to the phase of the oscillator, the amplitude has the chimeric feature as well. To see that clearly, we show in Fig. \ref{swc_amp} the snapshot of $|W|$ and $|S|$, i.e., the amplitude of $W$ and $S$. For $|W|$ in Fig. \ref{swc_amp}, it is found that $|W|$ is continuous and smooth outside the core region but randomized and discontinuous inside the core region. This can be further seen by plotting $\Delta |W|=|W_{i+1,N/2}|-|W_{i,N/2}|$ along the  center horizontal line as illustrated in Fig. \ref{swc_amp}(b). Similar to the phase in Fig. \ref{swc}, the fluctuation is quite big inside the core but almost vanishes, i.e., $\Delta |W| \approx 0$, outside the core region. Figure \ref{swc_amp}(c) shows the state of each oscillator in the phase portrait expanded by ${\rm Re}W$ and ${\rm Im}W$ for one moment (not a short period of evolution of the oscillator). One may note from Fig. 2(c) that the distribution of all the oscillators are far from the center and a hole exists there, which implies that $|W|$, the amplitude of the variable $W$, is quite large for each oscillator in this case. That is, the amplitude of the spiral pattern does not drop to zero even for the oscillator in the core region. This fact significantly differs from the coherent spiral waves for which the amplitude of the oscillator usually becomes smaller as it approach the core region, and various amplitudes should exist. This also implies that the two oscillator which stays very close in the physical space may be not close at all in the state space. In other words, the spatiotemporal patterns have the discontinuous features. Differing from the variable $|W|$, the amplitude of the diffusive environment variable $S$, i.e., $|S|$, shown in Fig. \ref{swc_amp} (d-e) shows continuous and smooth features. (The small gap between the left and right branch in Fig. \ref{swc_amp}(e) is due to the zero value of $|S|$ at the phase singularity). Significantly different from Fig. \ref{swc_amp}(c), there is no hole observed in the state space of ${\rm Re}S-{\rm Im}S$, which in turn implies that the resulted pattern of ${\rm Re}S$ or ${\rm Im}S$ is smooth.

\begin{figure}[t]
	\includegraphics[width=0.9\linewidth]{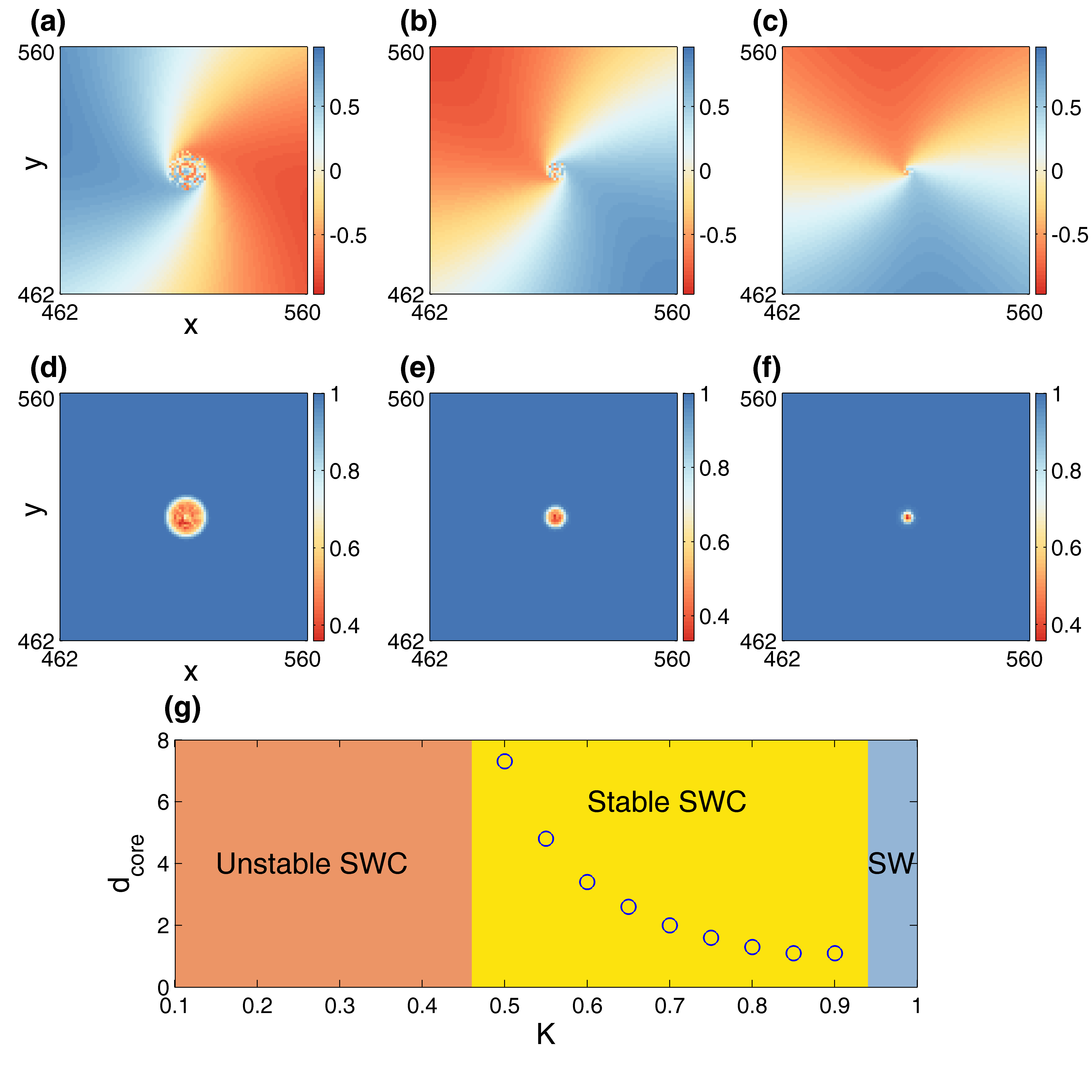}
	\caption{(color online). The dynamical state of spiral wave chimeras as a function of $K$. (a-c) three spiral wave chimeras for $K=0.6$, $0.7$ and $0.8$. (d-f) The averaged order parameter $\left<R\right>$ corresponding to (a-c). (g) Regions for different dynamical states for $K$. The circles in this panel denote the core diameter for the corresponding coupling strength $K$. SWC: spiral wave chimera. SW: spiral wave. The local dynamics parameter $\alpha$ is $-0.2$.}\label{dcore}
\end{figure}

\begin{figure}[t]
	\includegraphics[width=0.9\linewidth]{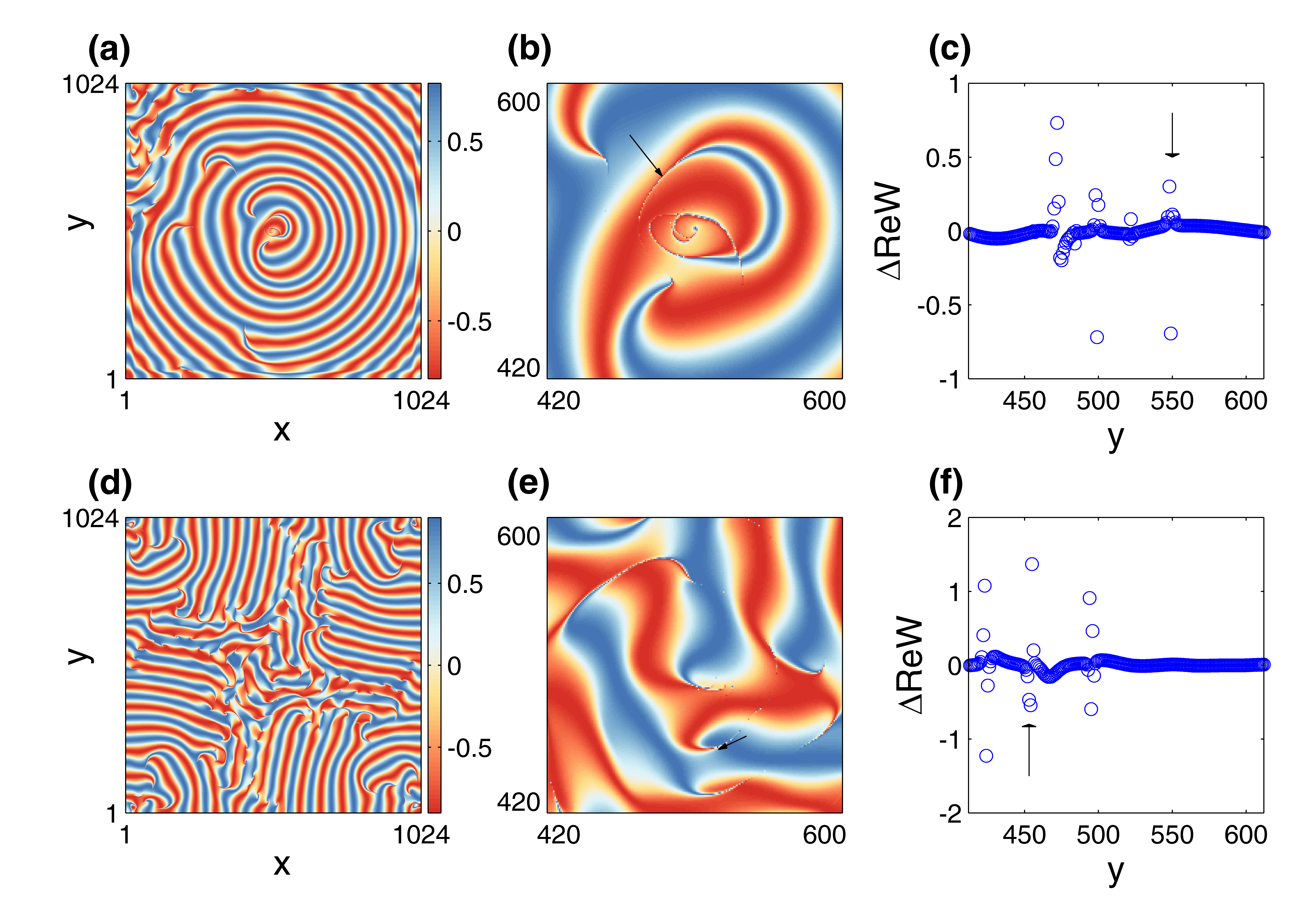}
	\caption{(color online). Two typical states of unstable spiral wave chimeras observed in the small coupling strength $K$. (a) Spiral wave chimera state with core break and (b) enlarged view of the core region in (a) for $K=0.35$. (c) The variation of $\Delta {\rm Re}W = {\rm Re}W_{491,j+1}-{\rm Re}W_{491,j}$ with respect to $y$ in (b). (d) A turbulent-like state and (e) enlarged view of the center region in (d) for $K=0.2$. (f) The variation of $\Delta {\rm Re}W = {\rm Re}W_{522,j+1}-{\rm Re}W_{522,j}$ with respect to $y$ in (e). Other parameters are the same as in Fig. \ref{dcore}.}\label{uswc}
\end{figure}

\begin{figure}[t]
	\includegraphics[width=\linewidth]{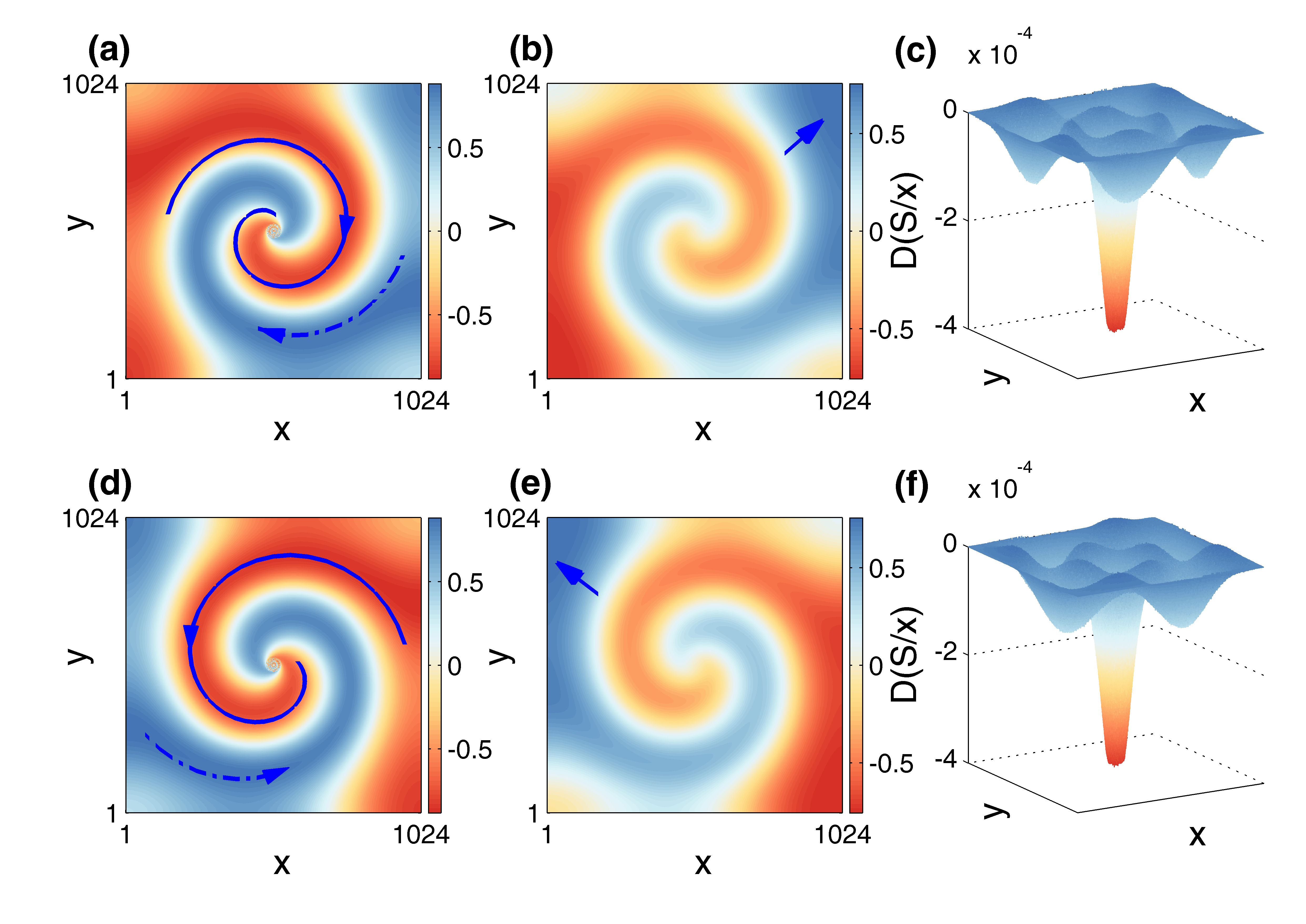}
	\caption{(color online). The effects of $\alpha$ on rotation direction of spiral wave chimeras for $K=0.65$. (a) A spiral wave chimera for ${\rm Re}W$ and (b) spiral wave for ${\rm Re}S$ rotating clockwise for $\alpha=+0.35$. (c) The spatial distribution of $D(S/x)$. (d) Spiral wave chimeras for ${\rm Re}W$ and (e) spiral wave for ${\rm Re}S$ rotating counterclockwise for $\alpha=-0.35$. (f) The spatial distribution of $D(S/x)$. The arrows with the solid (dashed) line denote the curl (rotation) direction in (a) and (d). The arrows in (b) and (e) denote the direction of wave propagation.}\label{rotation_rev}
\end{figure}

\subsection{The effects of coupling strength $K$ and local dynamic parameter $\alpha$}

In the preceding section, we have shown the existence of SWCs in the system described by Eqs. (\ref{SL1}-\ref{SL2}). For this system, the parameter $K$ determines the coupling strength between the oscillator and environment, and $\alpha$ represents the oscillatory frequency of the oscillator. To give more insights into the dynamics of SWCs in this system, we investigate how these parameters affect the behavior of SWCs in this section. At first, we identify the role played by the coupling strength in the dynamics of SWCs. To that, we keep $\alpha=-0.2$ and vary $K$.  Our simulations indicate that stable SWCs are observed in a wide range of coupling strength. Figure \ref{dcore}(a-f) shows SWCs and corresponding time averaged order parameter $\left<\text{R}\right>$ for three different coupling strengths $K=0.6$, $0.7$ and $0.8$. A quantitative dependence of $d_{core}$ on $K$ is plotted in Fig. \ref{dcore} (g). A clear fact is that the size of the incoherent core, say $d_{core}$, decreases as coupling strength $K$ increases. As $K$ approaches $K_{c} \approx 0.95$, $d_{core}\approx 5dx$, which is almost the same size as that of the spiral wave core in Fig. \ref{swc} (g). (Note that the core diameter of the coherent spiral for the environmental variable is independent of the coupling strength.) In this case, we say that a spiral wave rather than a SWC emerges.

Further simulations show that other dynamical states could also be observed as $K$ varies. Specifically, for too weak coupling $K<K'_{c}$ where $K'_{c} \approx 0.46$, the SWC is no longer stable and complex spatiotemporal patterns emerge given $\alpha=-0.2$. Figure \ref{uswc} display two typical dynamical states observed in the range of $K<K'_{c}$. In Fig. \ref{uswc}(a), we find the coexistence of coherent waves with several broken waves. Instead of the circular desynchronized core, each broken wave has a very slim line along which the oscillators are desynchronized (see the arrows in Fig. \ref{uswc}(b) and Fig. \ref{uswc}(c) for example). Further decreasing $K$, as we find, leads to more and more broken waves and the whole patterns seems more disordered as shown in Fig. \ref{uswc}(d-f). It is noted that in nonlocally coupled systems, waves with such desynchronized lines have been observed  \cite{kuramoto_ptps03}.   

The system parameter $\alpha$ which determines the oscillation frequency of the single oscillator would also influence the dynamics of SWCs. Two effects are observed in our simulations as changing $\alpha$. First, a transition from stable spiral wave chimera states to complex spatiotemporal patterns is observed. The resulted patterns are quite similar to those shown in Fig. \ref{uswc} caused by the change of $K$. Second, the resulted spatial patterns are almost unchanged if we only change the sign of $\alpha$, i.e., $\alpha \rightarrow -\alpha$. This is because for the isolated oscillator, changing the sign of $\alpha$ only alters the relative phase of ${\rm Re}W$ and ${\rm Im}W$. For a stable SWC, its rotation direction reverses for the same initial condition as $\alpha$ changes to $-\alpha$. For instance, we show in Fig. \ref{rotation_rev} two SWCs for $\alpha=+0.35$ and $\alpha=-0.35$, respectively. For $\alpha=+0.35$, a SWC rotates clockwise (CW) [see the dashed arrow in Fig. \ref{rotation_rev} (a)] and waves propagate outward [see the arrow in Fig. \ref{rotation_rev}(b)];  While for $\alpha=-0.35$, the SWC rotates counterclockwise (CCW)  [see the dashed arrow in Fig. \ref{rotation_rev} (d)] and waves still propagate outward [see the arrow in Fig. \ref{rotation_rev} (e)].  The panels (c) and (f) show the value of $D(S/x)$ whose definition is given by Eq. \eqref{DS} and physical meaning will be explained later.

\begin{figure}[t]
	\includegraphics[width=\linewidth]{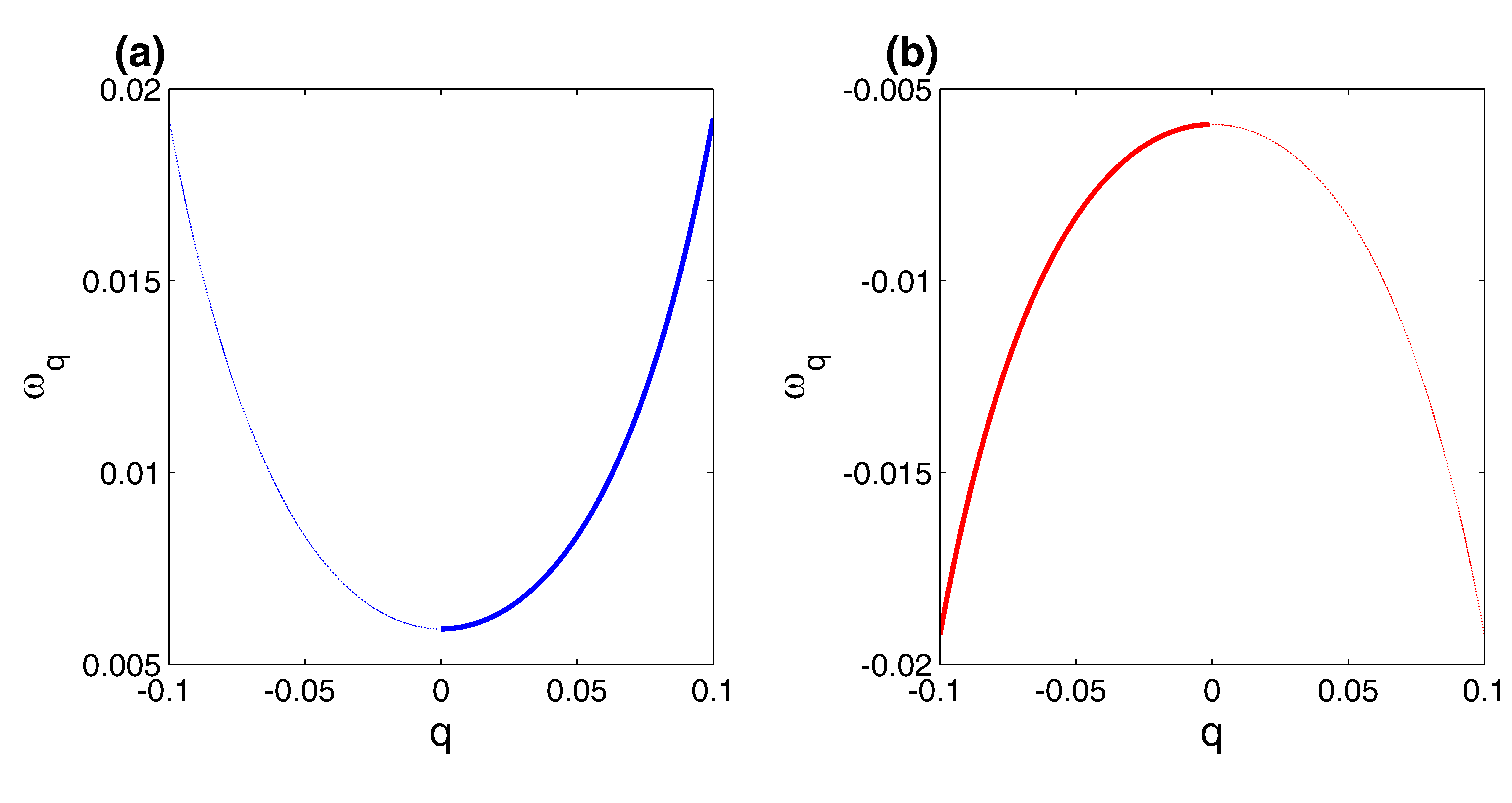}
	\caption{(color online). The dispersion relation given by Eq. \eqref{omega_q}. (a) $\alpha = 0.35$. (b) $\alpha=-0.35$. The other parameters are the same as in Fig. \ref{rotation_rev}.}\label{dispersion}
\end{figure}

The reversal of the rotation direction of the SWC as we change the sign of $\alpha$ is related to the direction of wave propagation and the conservation of topological charges. There are two kinds of velocity of the wave propagation, say group velocity $v_{gr}$ and phase velocity $v_{ph}$ which are defined as
\begin{eqnarray}
v_{gr} = \frac{\partial \omega_{q}}{\partial q},\quad v_{ph} = \frac{\omega_{q}}{q},
\end{eqnarray}
where $\omega_{q}$ and $q$ denote frequency and the wave number, respectively. The group velocity $v_{gr}$ determines the transport direction of the applied small perturbation, and usually for spiral waves it is always positive. However, the phase velocity $v_{ph}$ would either be positive (waves propagating outwardly) or negative (waves propagating inwardly). For SWCs, waves far way from the (desynchronized) center can be approximately viewed as the plan waves. Assuming the plan wave along the $x$ axis, we could write it in the following form 
\begin{eqnarray}
W = \rho_{w}e^{{\rm i}\left({\omega_{q}t-qx}\right)},\quad S = \rho_{s}e^{{\rm i}(\omega_{q}t-qx+\varphi)}. \label{velocity}
\end{eqnarray}
Substituting Eq.\eqref{velocity} into Eqs. \eqref{SL1} and \eqref{SL2}, we then get the following implicit dispersion relation,
\begin{eqnarray}
\omega_{q} = \alpha(1-K)+\frac{\alpha K(1+q^2\tau)}{(1+q^2\tau)^{2}+\omega_{q}^{2}}-\frac{K\tau\omega_{q}}{(1+q^2\tau)^{2}+\omega_{q}^{2}}. \label{omega_q}
\end{eqnarray} 
We show the dispersion relation in Fig. \ref{dispersion} for $\alpha=0.35$ and $\alpha=-0.35$. As required by $\partial \omega_{q}/\partial q>0$, the only right (left) branch could be possible for $\alpha=0.35(-0.35)$. For both cases, the phase velocity, $v_{ph}=\omega_{q}/q$, is always positive which means outward propagation of waves.

Another factor is the conservation of topological charges. Specifically, once the initial condition is given, topological charges should be conserved during the evolution if there are no topological defects disappearing through the boundary. As the environment variable $S$ is a spiral wave and its topological charge can be computed as \cite{pan_pre13}
\begin{eqnarray}
\sigma = {\rm sgn}[D(S/x)_{\rm PS}].
\end{eqnarray}
Here $D(S/x)$ is the $z$ component of $\nabla S_{1}\times \nabla S_{2}$ with $S_{1}= {\rm Re}S$ and $S_{2}={\rm Im}S$ being the real and the imaginary part of the complex
field $S$. That is,
\begin{eqnarray}
D(S/x) = \frac{\partial S_{1}}{\partial x}\frac{\partial S_{2}}{\partial y} - \frac{\partial S_{1}}{\partial y}\frac{\partial S_{2}}{\partial x}.\label{DS}
\end{eqnarray}
For a spiral wave, $D(S/x)$ takes the maximal or minimal value at the phase singularity. Figure \ref{rotation_rev} (c) and (f) shows $D(S/x)$ for two different cases and they are almost the same as we expect. According to Ref. \cite{pan_pre13}, for the same conditions, there are at most four different configurations of spiral waves for different parameters: two for outwardly propagating waves and the other two for inwardly propagating waves. Moreover, the curl direction $C$ [see the arrow with the solid line in Fig. \ref{rotation_rev} (a) and (d)], rotation direction $R$ and propagation direction $P$ satisfy following
relationship as found before \cite{pan_pre13}
\begin{eqnarray}
C\cdot R = P,
\end{eqnarray}
where $C=+1(-1)$ for CCW (CW), $R=+1(-1)$ for CCW (CW) and $P=+1(-1)$ for outward (inward) propagation. In our case, $P=+1$ for both $\alpha=0.35$ and $\alpha=-0.35$ as discussed above. Therefore, there are two possible configurations : one is $(C,R)=(-1,-1)$ and the other is $(C,R)=(+1,+1)$, which correspond to Fig. \ref{rotation_rev} (a) and (d), respectively.
\subsection{Estimation of $K_{c}$}
The critical coupling strength $K_{c}$ below which SWCs would arise could be estimated by the following way. As stated previously, the coupled equation represents a system among which oscillators are coupled indirectly through the diffusive environment $S$. The emergence of SWCs can be viewed a continuation problem from the spiral wave solution as $K$ changes. For $K>K_{c}$, the system admits the existence of a smooth core for both $W$ and $S$. It is known that for this system, the value of $W$ at the center (i.e., phase singularity) vanishes, i.e., $W(\mathbf{r}_{cent},t)=0$. Similarly, the value of $S$ at the spiral center is also vanished. As $K$ decreases to the critical value $K_{c}$, the oscillator at the spiral core center may lose its stability and become oscillatory, though the value of $S$ at the phase singularity still vanishes. Therefore, the onset of the SWC pattern can be regarded as the problem that the central oscillator becomes unstable for $K_{c}$ due to the interaction between the oscillator and environment. In other words, to estimate $K_{c}$, we need to check the stability of the central oscillator with the environmental coupling. The equation of the central oscillator reads,
\begin{eqnarray}
\frac{d W_{cent}}{dt} = W_{cent}-(1+i\alpha)|W_{cent}|^{2}W_{cent}-KW_{cent}. \label{W_center}
\end{eqnarray}
The above equation has a unique stationary solution $W^{ss}_{cent}=0$. Let $\delta W= W_{cent}-W^{ss}_{cent}$, we get the evolution of the perturbation $\delta W$ as,
\begin{eqnarray}
\delta \dot{W} = [(1-K)]\delta W
\end{eqnarray} 
which means the perturbation $\delta W$ behaves like
\begin{eqnarray}
\delta W \propto e^{(1-K)t}
\end{eqnarray}
It immediately concludes that the stationary solution will become unstable if 
\begin{eqnarray}
K<K_{c}=1. \label{kc}
\end{eqnarray}
This estimation is in agreement with our results as illustrated in the previous section where we find numerically the critical value of $K_{c}$ is about $0.95$. We note that $K_{c}$ estimated by Eq.\eqref{kc} is also true for the case of $\tau_{s}=0$ as considered in the previous work \cite{kuramoto_ptps03}.

\subsection{Analysis of SWCs formation}
\begin{figure}[t]
	\includegraphics[width=0.9\linewidth]{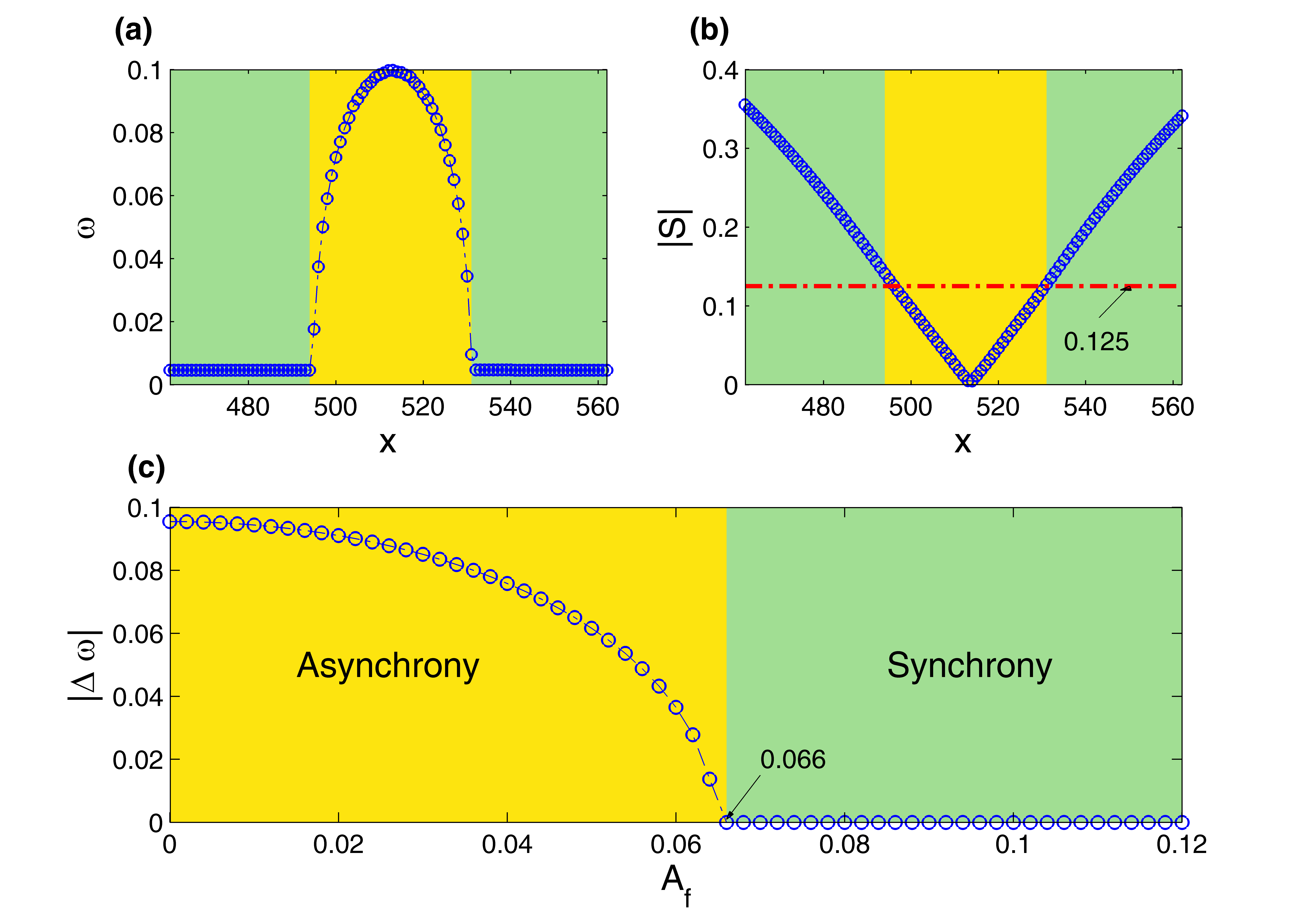}
	\caption{(color online). Mechanism analysis of the spiral wave chimera formation. (a) Frequency profile with respect to $x$ along the center line in Fig. \ref{swc}(a). The left and right light green regions mean synchronization between $W$ and $S$ while the centered yellow region denotes the desynchronized. (b) $|S|$ with respect to $x$ along the center line. (c) Frequency difference between the forcing and measured frequency, i.e., $|\Delta \omega|=|\omega_{m}-\omega_{f}|$ as a function of the forcing amplitude $A_{f}$ for a local system with $\omega_{f}=0.0046$. Other parameters are the same as in Fig. \ref{swc}.}\label{mechanism}
\end{figure}

Physically, the formation of SWCs in such a system can be further analyzed as follows. At first, the coupled system represented by Eqs. (\ref{SL1}-\ref{SL2}) can be viewed as the picture that the local oscillators are isolated from each other, but they are subjected to a spatiotemporal forcing from the environment variable $S$ in a self-organized manner except the most central oscillator. This central oscillator is absent from the forcing because of the vanished value of $S$ at the center. Consequently, the oscillation frequency can be computed analytically by substituting $W_{cent}=\rho \exp{(-{\rm i}\omega t})$ into Eq. (\ref{W_center}) and get $\omega = \alpha (1-K)$. For the present case of $K=0.5$ and $\alpha=-0.2$, we get $|\omega| = 0.1$ which is the same as that measured directly from the numerical simulations as illustrated in Fig. \ref{mechanism}(a). (Please refer to the frequency of the central oscillator.) 

While for other oscillators in the incoherent core, they are affected by the environment $S$ more or less. Further rewriting Eq. (\ref{SL1}) and denoting $F_{p}(t)\equiv KS_{p}(t)$ as the driving force from $S$ for the oscillator located at the position $p$, then its dynamics is governed by 
\begin{eqnarray}
\frac{dW_{p}}{dt} = (1-K)W_{p}-(1+i\alpha)|W_{p}|^{2}W_{p}+F_{p}(t).\label{period_forcing}
\end{eqnarray}
Please note that the driving force depends not only on time but also space. As $S_{p}(t)$ shows simple harmonic oscillation and we then approximately replace $F_{p}(t)$ by a periodic forcing as $F_{p}(t)=A_{f}\exp{({\rm i} \omega_{f}t)}$ where $A_{f}$ and $\omega_{f}$ are the amplitude and frequency of the forcing, respectively. With this approximation, the formation of SWCs can be analyzed by a phenomenological approach with the details below. 

Equation (\ref{period_forcing}) represents a classical situation: an oscillator subjected by a periodic forcing. According to the synchronization theory, an oscillator with the natural/intrinsic frequency $\omega_{0}$ could be locked to an external periodic driving only when the frequency mismatch between them is small and the amplitude is sufficient large. Or in the other words, for a given frequency mismatch, the synchronization between the oscillator and the external forcing occurs only if the amplitude larger than a critical value, i.e., $A_{f}\ge A_{f}^{c}$. 

For the case of Fig. \ref{swc} where $K=0.5$ and $\alpha=-0.2$, the natural frequency of the oscillator is $|\omega_{0}|= 0.10$. Except the phase singularity, the rotation frequency for $S$ is the same everywhere. Therefore, we take the driven frequency $\omega_{f}$ as the same as the rotation frequency of spiral waves of $S$, i.e., $\omega_{f}=\omega_{s}=0.0046$. With this setting, the frequency difference between the measured frequency $\omega_{m}$ and forcing frequency, i.e., $\left|\Delta \omega\right| = \left|\omega_{m}-\omega_{f}\right|$, as a function of $A_{f}$ is shown in the Fig. \ref{mechanism} (c). Evidently, only for $A_{f}>A_{f}^{c}$, $\Delta \omega$ tends to be zero which means synchronization between the oscillator and external forcing occurs. For the current case, we find $A_{f}^{c} = 0.066$. This suggests that the critical amplitude of $S$ should be $A_{s}^{c} = A_{f}^{c}/K=0.132$. 

On the other hand, the yellow center region shown in Fig. \ref{mechanism}(b), which is the same as the one in Fig. \ref{mechanism}(a), denotes the desynchronized region in which the oscillators can not be synchronized because of the too weak forcing from the environmental variable $S$. The left (right) boundary of this region interacts with the left (right) branch of the curve (see blue circles) showing the dependence of $|S|$ on the space. We find the value of $|S|$ at the intersection point is approximately $0.125$. Beyond these boundaries, the amplitude of $S$ such as in the green region is larger than this critical value and then oscillators can be forced to be synchronized. We find that this critical amplitude of $S$ (i.e., 0.125) is close to the predicted value $A_{s}^{c}=0.132$ from the synchronized theory.

\section{Spiral wave chimeras in a FitzHugh-Nagumo oscillator system}
\subsection{The three-component FitzHugh-Nagumo model}
The existence of spiral wave chimeras in populations of oscillators coupled via a slowly changing environment, as we find, is quite robust and not dependent on the specific model. To show that, we choose another kind of classical oscillator such as the FHN type oscillator as the local dynamics of the system. Specifically, the intrinsic dynamics variables are chosen as $\mathbf{Z}=(u,v)$ and the dynamical functions as $\mathbf{F}(\mathbf{Z})=(au-\gamma u^3-v,u-v)$, and we set the environmental variable $\mathbf{z}=w$. Then the interaction terms are chosen as  $\mathbf{H}(\mathbf{Z},\mathbf{z})=-\eta w$ and $\mathbf{G}(\mathbf{Z},\mathbf{z})=u-w$. Finally, the coupled equations read \cite{alonso_jcp11,li_pre16,li_cnns21}
\begin{eqnarray}
\frac{\partial u}{\partial t} &=& au -\gamma u^{3}-bv -\eta w,\label{rd1}\\
\frac{\partial v}{\partial t} &=& u - v, \label{rd2} \\
\tau_{w}\frac{\partial w}{\partial t} &=& u - w + D_{w}\nabla^{2} w. \label{rd3}
\end{eqnarray}
Here $u$ and $v$ represent an activator and inhibitor, respectively. They denote the state of the oscillator. The variable $w$ can be viewed as the external environment through which oscillators communicate with each other. $a$, $\gamma$ and $b$ are the parameters determining the intrinsic dynamics of the oscillator. $\eta$ is related to the coupling strength and $D_{w}$ is the diffusion coefficient of the environment variable. $\tau_{w}$ is the characteristic time of $w$. It is noted that we have shown previously that SWCs exist in a wide parameter regime of this system and even breakup of SWCs \cite{li_pre16,li_cnns21}. However, our previous works mainly focus on a finite value of $\tau_{w}$ which is not large. Here we just further show an example of that SWCs exist even when $\tau_{w}$ is large enough, similar to the case of SL oscillators.

We employ the explicit Euler-forward method to solve these coupled equations with a space step $dx=dy=0.1$ and a time step $dt=(dx)^2/5=0.002$. As we consider the situation of the slowly changing field, we set $\tau_{w}=100$ too. The effective diffusion constant $D_{\rm eff} = D_{w}/\tau_{w} =1$ for simplicity as previous. The other parameters such as $a=4.0$, $\gamma=4/3$, $b=2.5$, $\eta=3.5$ are fixed.

\begin{figure}[t]
	\includegraphics[width=\linewidth]{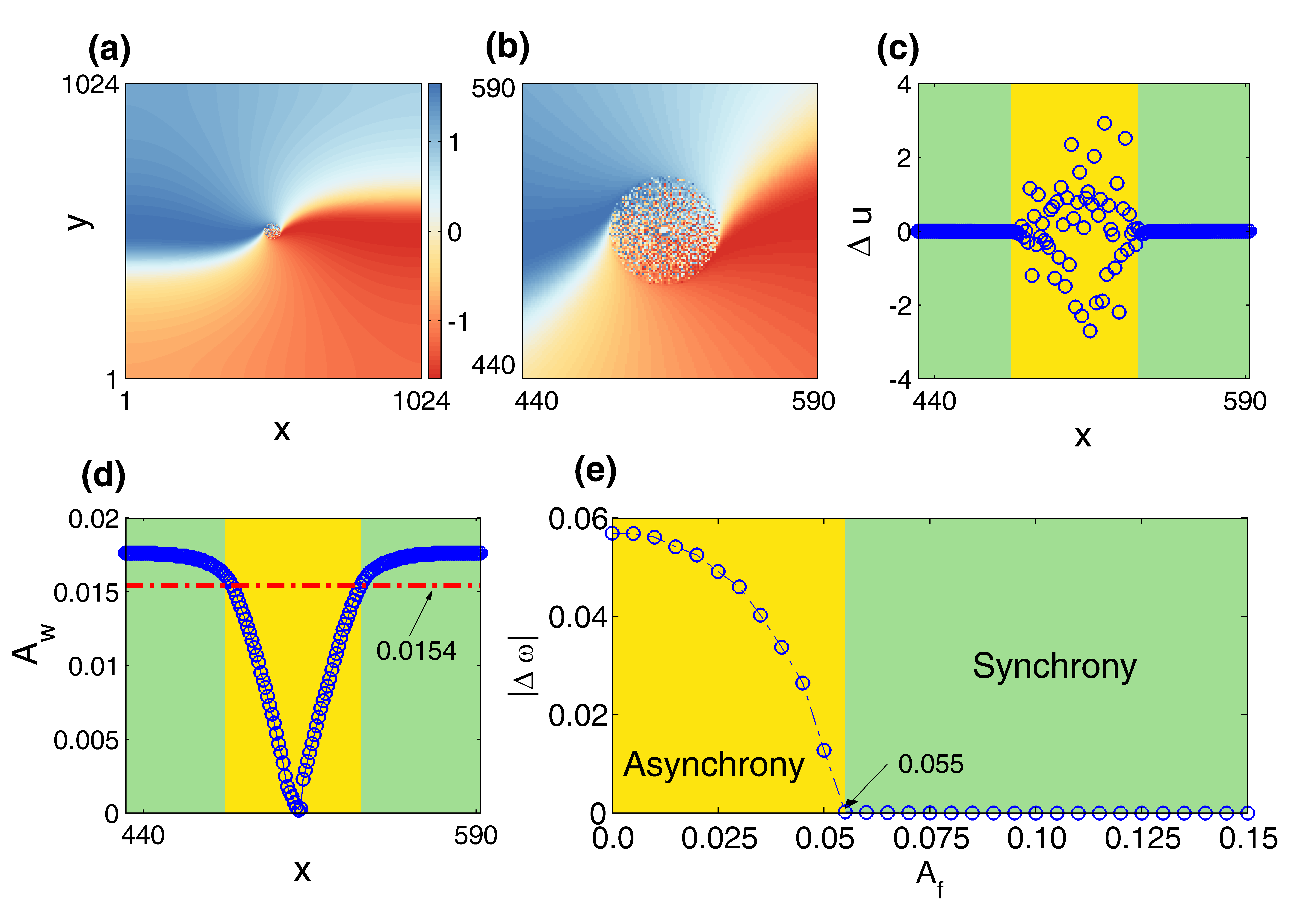}
	\caption{(color online). Spiral wave chimeras in populations of FHN oscillators coupled via a slowly changing environment. (a) A snapshot of spiral wave chimeras for the $u$ component and (b) enlarged view of the core region in (a). (c) The variation of $\Delta u=u_{i+1,N/2}-u_{i,N/2}$ with respect to $x$ along the center line. (c) Time-averaged amplitude, $A_{w}(x)$, along central horizontal axis in (b). Frequency difference between the forcing and measured frequency, i.e., $|\Delta \omega|=|\omega_{m}-\omega_{f}|$ as a function of forcing amplitude $A_{f}$ for a local system.}\label{swc_fhn}
\end{figure}
\subsection{Spiral wave chimera states in FHN system}
The existence of a spiral wave chimera in populations of FHN oscillators coupled via a slowly varying environment is shown in Fig. \ref{swc_fhn}. Figure \ref{swc_fhn}(a) and (b) display a spiral wave chimera state and its enlarged view of the core center. The rotation frequency of the spiral arm is $\omega_{s}=0.870$. The difference between the adjacent oscillators $\Delta u=u_{i+1,N/2}-u_{i,N/2}$ is illustrated in Fig. \ref{swc_fhn}(c) where the desynchronized region is highlighted by the yellow shaded region. Similar to the observation in the SL system, we find that the nondiffusive component $u$ (or $v$) shows the feature of spiral wave chimeras and the diffusive variable $w$ demonstrates the spiral wave with a smooth core. This facts together with previous findings strongly suggest that spiral wave chimera states in an ensemble of oscillators coupled via a slowly varying of the environment are model independent.

The underlying mechanism of the occurrence of spiral wave chimeras in the FHN coupled system can also be analyzed from the point of synchronization driven by external forcing. That is, the FHN oscillators ($u,v$) are subjected to a spatiotemporal forcing from the field of $w$. Inside the core region, the amplitude of $W$ denoted by $A_{w}$ is too weak to synchronize the FHN oscillators; while outside the core, $A_{w}$ is strong enough to synchronize the FHN oscillators. To illustrate that, we plot the amplitude $A_{w}$ as a function of position along the center line. It is evident that inside the core, the amplitude drops to a smaller value (shaded region). According to the core size, we find in this case $A_{w}$ should be larger than $0.0154$. This critical value can be predicted by the synchronization of oscillators driven by an external forcing. Specifically, as the time evolution of the component $w$ like a harmonic oscillation, we then replace $-\eta w$ by $A_{f}\sin(\omega_{f}t)$, i.e.,
\begin{eqnarray}
\frac{du}{dt} &=& au-\alpha u^{3}-bv+A_{f}\sin(\omega_{f}t),\\
\frac{dv}{dt} &=& u-v.
\end{eqnarray}
For a single FHN oscillator, we get the frequency $\omega_{0}=0.812$. The driven frequency $\omega_{f}$ is chosen as the frequency of $w$ and in the present case $\omega_{f}=\omega_{s}=0.870$. The frequency difference $\left|\Delta \omega\right| = \left|\omega_{f}-\omega_{m}\right|$ where $\omega_{m}$ is the frequency of the oscillator responded to the periodically driven as a function of $A_{f}$ is illustrated in Fig. \ref{swc_fhn}(e). We find that the FHN oscillator can be synchronized driven by the period forcing only if $A_{f}\ge A_{f}^{c}=0.055$. This means that, the critical value of amplitude $A_{w}^{c}$ for $w$ should be $A_{w}^{c} = A_{f}^{c}/\eta = 0.0157$, which is well in agreement with the value $0.0154$ as indicated by the dot-dashed line in Fig. \ref{swc_fhn}(d). 

\section{Discussion and conclusion}

The environmental coupling is a quite common coupling mechanism for observations of synchrony and pattern formation in a diversity of systems ranging from physical to chemical and biological systems. In this work, we have systematically investigated the dynamics of SWCs in ensembles of oscillators coupled through a slowly changing diffusive environment. Our findings presented in the current work differ from the previous works in two aspects. Firstly, in the current work, the time scale of the dynamical environment is much larger than that of the intrinsic dynamics, i.e., $\tau \gg 1$, while in previous works the time scale $\tau$ is either $\tau \ll 1$ or close to one. For example, in the seminal work by Kuramoto {\it et al}., a key assumption to observe SWCs is that the time scale $\tau$ of the environmental variable is so small (i.e., $\tau\rightarrow 0$ ) that it can be eliminated adiabatically \cite{kuramoto_ptps03,shima_pre04}. Consequently, the original system is reduced to a nonlocal coupled oscillator system. On the other hand, SWCs have been shown to exist in the three-component FHN model for which the time scale of the environmental component is close to one \cite{li_pre16,li_cnns21}. Secondly, our main results are demonstrated with the intrinic dynamics described by environmentally coupled SL equation. This SL equation represents the normal form of a class of systems closed to Hopf bifurcation. Therefore, compared to the specific FHN model used in the previous works, this model represents a more general case, and then justifies the generality of the observation of chimera states in OEC systems.

In the current work, SWCs would be unstable as we decrease the coupling strength $K$. The main difference between two unstable states as we observe is that the one in Fig. 4(d) for $K=0.20$ seems more disordered than that in Fig. 4(a) for $K=0.35$. According to the evolution of the state with $K=0.35$, the initial small SWC at the center of the system as we observe expands at first and try to organize the whole medium (though it fails finally); While for $K=0.20$ the initial small SWC shrinks and finally almost disappear. These facts imply that the latter is more unstable than the former one. The exact mechanism underlying the instability of the SWCs seems complicated and remains unclear now. Exploring the instability mechanism is beyond the scope of this work, which we leave for future work.

For experimentalists, there is a key question that which kinds of coupled systems are feasible to be realized in experiments. The present findings may provide some hints to this issue. Together with previous work \cite{tang_jcp14,li_pre16,li_cnns21}, it shows that emergence of SWCs is quite robust in the system representing a large population of oscillators coupled via a diffusive field. There is a broad class of the systems with similar environmental coupling schemes such as chemical oscillators BZ particles immersed in catalyst-free solutions \cite{taylor_sci09}, social amoeba {\it Dictyostelium discoideum} \cite{noorbakhsh_pre15}, genetically engineered bacteria \cite{danino_nat10}, yeast cells \cite{shutz_BJ11}. Therefore, we expect that chimera states are highly possible in these biological or chemical systems. On the other hand, the realization of the nonlocal coupling in experiments up to date still strongly relies on a computer algorithm and the used system is often discretized. A main challenge in such experimental settings is to overcome the oscillator number limit. For example, realization of $10,000$ chemical oscillators with the nonlocal coupling with the help of computer seems a quite difficult task in the experiment with chemical oscillators \cite{totz_phdtheis}. Therefore, it is not easy to extend such a two dimensional system directly to a three dimensional one where the oscillators involves much more. However, for the system presented here, the coupling is mediated by diffusion which occurs in a more natural way, and it may be easily extended to large systems and to investigate the dynamics of scroll wave chimeras in three dimension once it is realized experimentally.

In summary, by coupling a large group of oscillators to a slowly changing diffusive environment, we have shown numerically that SWCs do exist in such systems. Given $\alpha$, to observe the SWCs, the coupling strength could not be too large or too small. As $K$ is increased, a transition from SWCs to spiral waves with smooth cores could occur; while $K$ decreases and is lower than a certain critical value, SWCs will be unstable and new kind of chimera structures could emerge. The change of $\alpha$ would also cause the instability and reverse the rotation of spiral wave chimeras. We further confirm that spiral wave chimera states exist in FHN systems with the similar coupling scheme, which in turn suggests the robustness of our findings. As our systems are in analogy to some biological and chemical systems, the findings in the present work provide hints to generate chimera states in realistic systems.\\

{\noindent \bf ACKNOWLEDGMENTS} 

This work was supported by the National Natural Science Foundation of China under Grants No. 11875120 and Natural Science Foundation of Zhejiang Province under Grant No. LY16A050003.

\end{document}